\documentclass[12pt,preprint]{aastex}

\def\mearth{ M_{\earth}}

\def\rjup{ R_{J}}

\def\rsun{\rm R_{\odot}}
\def\rs{R_{\star}}
\def\ms{M_{\star}}
\def\msun{\rm M_{\odot}}
\def\fdg{\mbox{\ensuremath{.\!\!^\circ}}}

\begin{document}

\title{A Precise Estimate of the Radius of the Exoplanet HD~149026b from \emph{Spitzer} Photometry}

\author{	Philip Nutzman		\altaffilmark{1},
		David Charbonneau 	\altaffilmark{1,2},
		Joshua N. Winn		\altaffilmark{3},
		Heather A. Knutson		\altaffilmark{1},
		Jonathan J. Fortney	\altaffilmark{4},
		Matthew J. Holman		\altaffilmark{1},
		Eric Agol		\altaffilmark{5}	 
		}

\altaffiltext{1}{Harvard-Smithsonian Center for Astrophysics, 60 Garden St., 
Cambridge, MA 02138}
\altaffiltext{2}{Alfred P. Sloan Research Fellow}
\altaffiltext{3}{Department of Physics, and Kavli Institute for
Astrophysics and Space Research, Massachusetts Institute of
Technology, Cambridge, MA 02139, USA}
\altaffiltext{4}{Department of Astronomy and Astrophysics, UCO/Lick Observatory, University of California, Santa Cruz, CA 95064}
\altaffiltext{5}{Department of Astronomy, University of Washington, Box 351580, Seattle, WA 98195}

\email{pnutzman@cfa.harvard.edu}

\begin{abstract}

We present \emph{Spitzer} 8 $\mu$m transit observations of the extrasolar planet HD 149026b.  At this wavelength, transit light curves are weakly affected by stellar limb-darkening, allowing for a simpler and more accurate determination of planetary parameters.  We measure a planet-star radius ratio of $R_p/\rs = 0.05158 \pm 0.00077$, and in combination with ground-based data and independent constraints on the stellar mass and radius, we derive an orbital inclination of $i = 85 \fdg 4 ~^{+0 \fdg9}_{-0 \fdg8} $ and a planet radius of $R_p = 0.755 \pm 0.040 ~\rjup$.  These measurements further support models in which the planet is greatly enriched in heavy elements.

\end{abstract}

\section{Introduction}

Much attention has been lavished on the transiting extrasolar planet HD 149026b \citep{sato2005} due to its potential to directly test models of planet formation.  The planet's small observed radius for its mass imply that an extraordinary fraction of its mass (roughly 2/3) is in the form of heavy elements \citep{sato2005,fortney2006,burrows2007}.  The discovery of a metal-laden planet orbiting a very metal-rich host star ([Fe/H] $= 0.36$; \citealt{sato2005}) strongly suggests that core-accretion (e.g., Pollack et al. 1996) \nocite{pollack1996} plays a role in forming giant planets.  If, however, most of the heavy elements reside in the planet's core, then HD 149026b would possess a core mass much greater than the expected critical core mass of $10-20 \mearth$ \citep{mizuno1980,pollack1996}, and thus nonetheless present a challenge to standard core-accretion theory.  

The planet is noteworthy in another respect.  Observations by the \emph{Spitzer Space Telescope} \citep{harrington2007} have shown the planet to have a day-side 8 $\mu$m brightness temperature well in excess of its predicted blackbody temperature, when it is assumed that all incident radiation is absorbed and subsequently re-emitted uniformly across the entire surface of the planet.  \citet{fortney2008} posit that highly irradiated planets such as HD 149026b, which they term ``pM'' class planets, will generally show bright day-sides and large day/night temperature contrasts.  They argue that the incident stellar flux is prominently absorbed by gaseous TiO and VO high in the atmospheres of pM planets where the radiative timescale is much shorter than the advective timescale \citep[see also][]{hubeny2003,burrows2008}.  This is in contrast to less irradiated ``pL'' planets where Ti and V are expected to largely condense out of the atmosphere, permitting the stellar flux to be absorbed deeper in the atmosphere where the two timescales are comparable.  Hence, it is only for the pL class that a heated parcel of gas is able to be advected to the night side prior to cooling, resulting in similar day/night temperatures. 

HD 149026b is thus a valuable case study for modelers of planetary atmospheres, structure, and formation.  Unfortunately, the system is observationally challenging: the transit depth (3 mmag in $V$) is a factor of two shallower than any other presently known transiting planet, and more importantly, there are few adequate comparison stars nearby on the sky.  The result is that the present fractional uncertainty in the key observable parameter, the planetary radius $R_p$, is $7 \%$ \citep{winn2008}.  This uncertainty is one of the largest among the ensemble of transiting planets.  The state of uncertainty is unfortunate given that $R_p$ is the essential constraint on models of the planet's interior structure.  Fortunately there is further scope for improvement through high-precision photometry.

This study is inspired by the potential of infrared photometry with the \emph{Spitzer Space Telescope} to reduce the uncertainty in $R_p$.  While ground-based photometry suffers from significant levels of systematic noise when there are few good comparison stars, \emph{Spitzer} has demonstrated 0.1 mmag photometry without any comparison stars (e.g. Knutson et al. 2007)\nocite{knutson2007}.  Additionally, because of the near absence of stellar-limb darkening in the infrared, transit light curve modeling is simplified and gives results largely independent of assumptions about limb-darkening coefficients.  Previously, \citet{sato2005}, \citet{charbonneau2006}, and \citet{winn2008} (hereafter W08) have presented ground-based photometry of HD 149026. In this paper, we report \emph{Spitzer} 8 $\mu$m  observations of the transit of HD 149026b, and combine this with the previously published data in order to derive precise constraints on the properties of HD 149026b.  In \S 2 we describe the observations and data reduction and in \S 3 we describe our analysis of the \emph{Spitzer} light curve. In \S 4 we estimate the physical parameters of the HD 149026 system.  We conclude with a discussion of the implications of our revised estimate of the planet radius for models of the interior structure of HD 149026b.

\section{Observations and Reduction}
We observed the transit of HD~149026 on UT 2007 August 14, using the 8 micron channel of the IRAC instrument (Fazio et al 2004) aboard the \emph{Spitzer} Space Telescope \citep{werner2004}.  The system was observed at a 0.4 s cadence using IRAC's 32 by 32 pixel sub-array mode, in which frames of 64 images are taken in rapid succession.  Over the course of our observations, we obtained 1047 such frames, resulting in 67,008 total images.  Our observational strategy matches that of recent Spitzer observations of HD 189733 and GJ 436 (e.g., Knutson et al. 2007, Deming et al. 2007, Gillon et al. 2007); the telescope positioning was held fixed to avoid time loss during telescope movements and to minimize errors from an imperfect flat-field correction.  In the IRAC 8 micron channel, there is a well-known rise in detector sensitivity during observational sequences (see e.g., Harrington et al. 2007, Knutson et al. 2007), which is steepest at the beginning of observations and asymptotes within several hours for highly illuminated ($>$ 250 MJy Sr$^{-1}$) pixels.  We padded the beginning of our observational window so that the transit would begin nearly 3 hours into observations, thus avoiding the steepest part of this ``ramp.''

In each image, we assessed the background flux by taking the median pixel value from the corner regions of each 32 by 32 image. We performed aperture photometry, settling on a 3.5 pixel aperture radius, for which the rms of the time series is minimized.  From the time stamp reported for each frame of 64 0.4 s exposures, we calculated the JD of the center of integration for each image.  We applied the heliocentric correction to the JD using the position of \emph{Spitzer} obtained from the JPL Horizons Ephemeris System.  In each series of 64 images there is a well-known effect, with the first 5-10 and $58^{th}$ images showing anomalously low star fluxes and background levels (see e.g. Harrington 2007). Background subtraction generally corrects for this effect, but we elected to drop the 1$^{st}$ image from each series of 64, because the background levels in these images exhibit more dispersion than in the other images.  We trimmed the first 45 minutes of data, when the ramp is steepest.  We flagged images when the star centroid, calculated with a flux-weighted average, was 4$\sigma$ away from the median centroid position.  Such 4$\sigma$ centroid deviants were generally caused by cosmic rays or other contamination in the photometric aperture. We further flagged images when the flux measurement was 4$\sigma$ from a smoothed (binned) light curve, or the background level was 4$\sigma$ from a binned time series of the background.  We flagged 317 images (0.5 $\%$ of the total) according to the last three criteria.

\section{\emph{Spitzer} Light Curve Analysis}

One major benefit of observing transits at 8 $\mu$m is that stellar limb-darkening has a small effect on the shape of the transit light curve.  To determine its extent, we consulted a theoretical limb-darkening model \citep{kurucz1979,kurucz1994} for a $T_{\mathrm{eff}}= 6250 K$, $\log g = 4.5$, [Fe/H] $= 0.3$ star at $\lambda = 8 ~\mu$m.  We fit this model\footnote{See {\tt http://kurucz.harvard.edu/grids/}} to the \citet{claret2000} four parameter nonlinear limb-darkening law (see also Beaulieu et al. 2008 \nocite{beaulieu2008} for a similar handling of limb-darkening).  Though the limb-darkening is indeed modest, we find that incorporating it in our light curve modeling (described below) leads to non-negligible changes in the best-fit parameters and a reduction in the best-fit $\chi^2$ by more than 1.  We modeled the light curve using the ``small-planet" transit routine of Mandel \& Agol (2002) \nocite{ma2002}.  The small planet approximation is not usually suitable for analyzing high quality transit data, especially for systems with large planet-star radius ratios ($R_p/\rs \gtrsim 0.1$), but here we find the approximation leads to insignificant changes in the best-fit parameters (due to the very small planet-star radius ratio of HD 149026).  We assumed a circular orbit, which is expected from tidal dissipation and supported by current radial velocity data (e.g., Sato et al. 2005).  We parametrized the light curve with 4 geometric parameters that are independent of prior assumptions on the stellar properties: the planet-star radius ratio $R_p/\rs$, the stellar radius to orbital radius ratio $\rs/a$, the inclination $i$, and the time of mid-transit $T_c$.  To correct for the ramp and other possible detector effects, we adopted a correction factor $f = (c_{0} + c_{1} \log(t-t_0) + c_2 \log^2(t-t_0))$, where $t_0$ was fixed to a time a few minutes before the first observations.  Note that in all of our modeling below, we fit for the detector correction coefficients simultaneously with the transit-related parameters, allowing us to take into account how changes in the correction coefficients may impact the transit parameters.

\begin{figure}[ct1]
\epsscale{1.0}
\plotone{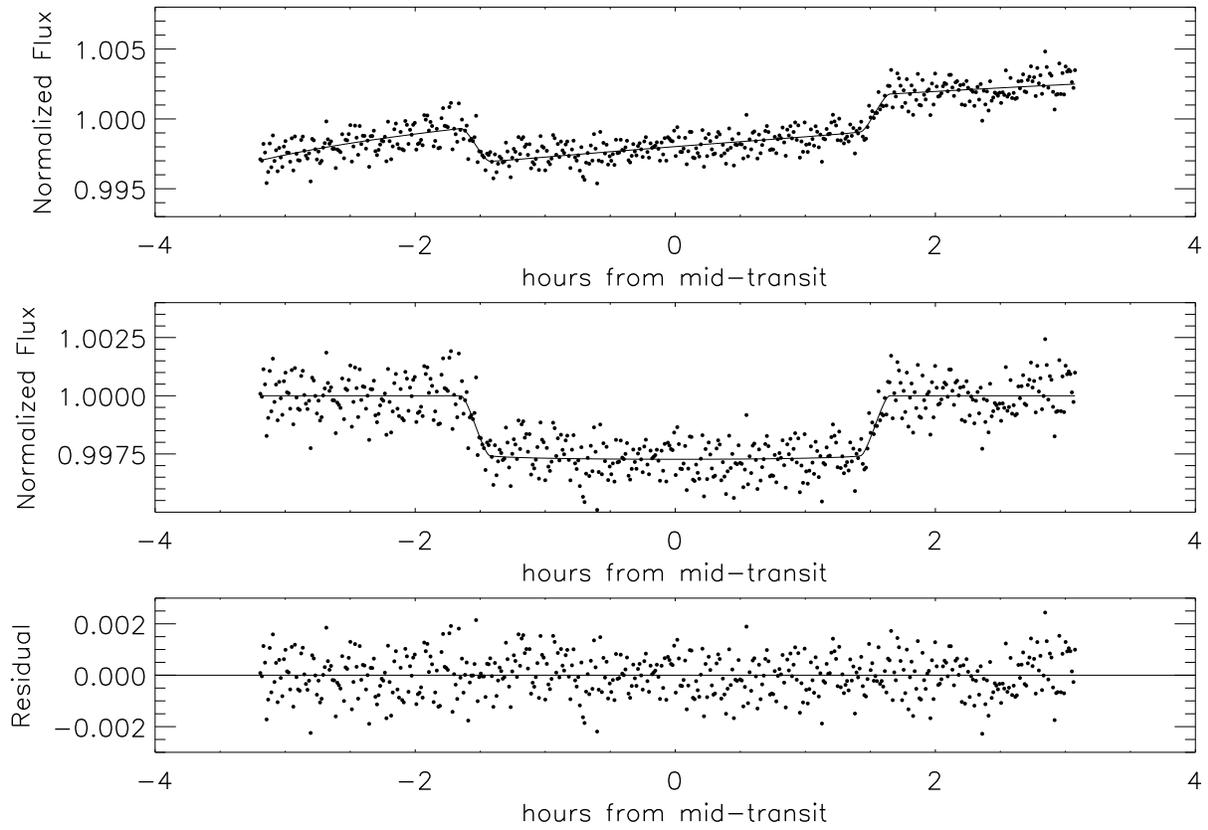}
\caption{Transit photometry for HD 149026, with 40 second resolution (bins of 100 images).  The top panel displays the raw light curve and the middle displays the light curve corrected for the detector ramp as described in section 3.  At bottom are the residuals from the best fit light curve.}
\label{fit}
\end{figure}

\begin{figure} 
\epsscale{1.0}
\plotone{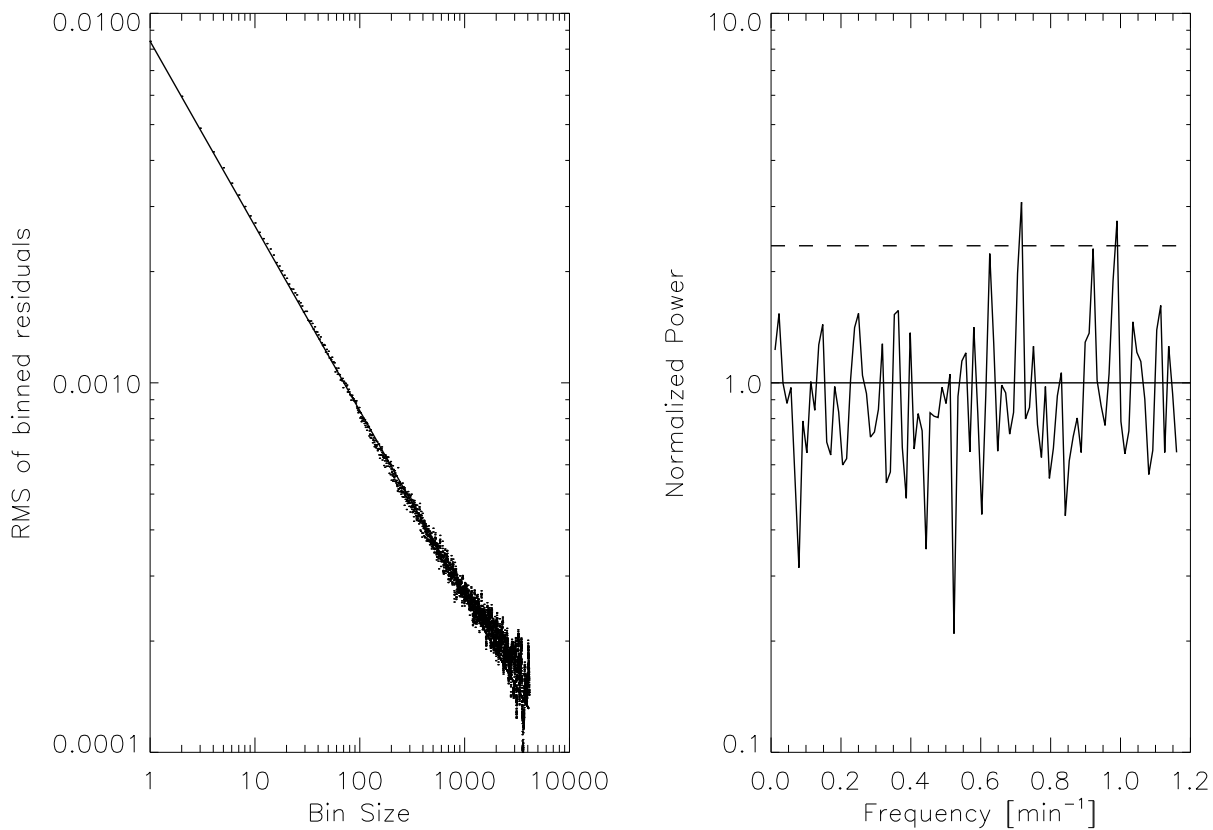}
\caption{Left: Root-mean-square of binned residuals vs. bin size.  The solid line is proportional to $N^{-1/2}$ and is normalized to match the value for bin size $N=1$.  Right: Power spectrum estimate for the time-series residuals.  The estimate has been divided by the power spectrum expected for randomly generated white noise of the same standard deviation and time sampling as the \emph{Spitzer} data.  The dashed line represents the median peak value of the simulated power spectrum estimates. }
\label{residual}
\end{figure}

We performed a least-squares fit to our unbinned data over the 7 parameter space ($R_p/\rs$, $\rs/a$, $i$, $T_c$, $c_0$, $c_1$, $c_2$), using an IDL implementation of the \texttt{amoeba} algorithm (e.g. see \citealt{press92}).  The data, corrected for the ramp and binned 100:1 are shown in Figure \ref{fit}, together with this best-fitting solution.
To understand the level of photometric noise and its properties, we examined the residuals from this best fit.  We determined a normalized rms residual of $8.3 \times 10^{-3}$, only 15\% greater than the expected photon-noise.  We found that the level of photometric noise was constant over the duration of the observations, and furthermore that the noise was essentially ``white.''  In the left panel of figure \ref{residual} we show that the scatter in binned residuals decreases with bin size as $N^{-1/2}$ for bins of up to 1000 images.  
In the right panel of Fig. 2, we display a power spectrum estimate for the time-series of residuals.  To compute this, we first binned the residuals for each frame of 64 images.  This step creates an evenly spaced time-series of 940 residuals because IRAC, in sub-array mode, takes exposures in sets of 64 images (once every 25.6 seconds).  This binning also avoids having to interpolate over gaps caused by flagged images.  Though binning removes the highest frequency information from the spectrum, we are less concerned with noise power on the affected time scales, which are shorter than the other timescales relevant in a transit light curve (e.g., the ingress/egress duration).  We estimated the power spectrum via the (modulus squared) discrete Fourier transform.  To reduce the variance at each frequency, we smoothed with a 7 point ``Daniell'', or moving average, filter.  We compared our power spectrum estimate with that expected of white noise by simulating $10^5$ time-series with identical, independent Gaussian deviates of the same time sampling and standard deviation as the \emph{Spitzer} residuals.  We note that $5\%$ of the simulated power spectra show peak values as high as the peak value in \emph{Spitzer} power spectrum, while two peaks in the \emph{Spitzer} power spectrum exceed the median peak value of the simulated spectra (the dashed line in Fig. 2).

\clearpage

\clearpage
 
Because of the light curve modeling degeneracy between the parameters $a$ and $\rs$, transit photometry alone cannot determine the quantity of interest, $R_p$.  To break this degeneracy, one can either apply an external constraint on $a$ (typically via Newton's version of Kepler's Third Law and a constraint on $M_\star$), on $\rs$, or on some combination of both.  Before applying any such constraints, we estimated the probability distributions for the 7 light curve parameters by using the widely employed Markov Chain Monte Carlo (MCMC) technique (see e.g., Tegmark et al. 2004; Winn et al. 2007; Burke et al. 2007).  The benefit of performing this analysis without any \emph{a priori} assumptions on stellar quantities is that we can compare, on equal footing, the light curve constraints derived from our data with constraints derived from other photometric data. For this analysis, we adopted a conventional $\chi^2$ function as our goodness-of-fit statistic:
\begin{equation}
\chi^2 = \sum_{i}{ \left(
\frac{f_{mod}(i)-f_{obs}(i)}{\sigma}\right)^2 } 
\end{equation}
where $f_{mod}(i)$ the calculated flux at the time of the i$^{th}$ data point, $f_{obs}(i)$ is the i$^{th}$ flux measurement, and $\sigma$ is fixed to the value of the rms determined in the previous paragraph.   We produced 10 chains of length $10^6$, with each chain starting from independent parameter points randomly chosen from a broad region (spanning approximately $5\sigma$) in parameter space.  The beginning $25\%$ of each chain was trimmed and the 10 chains were concatenated.  We found that the Gelman-Rubin R statistic was $< 1.01$ for each parameter, which is an indication of convergence.  

Our MCMC analysis yields $R_p/\rs = 0.05158 \pm 0.00077$ and impact parameter $|b| \equiv |a \cos i / \rs| = 0.62 ~^{+0.08}_{-0.24}$.  Our result for $R_p/\rs$ is larger than that of W08 ($R_p/\rs = 0.0491 ~^{+0.0018}_{-0.0005}$), though the difference is only modestly significant.  Our result for $|b|$ is higher than that of W08 ($|b|=0.00$ with 68\% upper limit $|b|=0.36$). 
We expect that our $R_p/\rs$ result is more robust than the result from the optical data because, for weakly limb-darkened IR light curves, the radius ratio is measured almost directly from the observed flux decrement.  For optical light curves, however, the radius ratio is strongly covariant with the assumed limb-darkening coefficients.  That is, an error in the assumed limb-darkening can translate into an error in the radius ratio.  To investigate this point, we conducted a comparison with previously published $(b + y)/2$ HD 149026 light curves of Sato et al. (2005) and W08  (see \S 4.2 for further discussion of these data).  For this sub-study we modeled the data in a manner similar to the above, but assumed a linear limb-darkening coefficient, which we allowed to vary freely.  For the optical light curves, we determined a correlation coefficient between $R_p/\rs$ and the limb-darkening coefficient of $r =0.55 $, while for the same experiment with the \emph{Spitzer} data we found a much weaker correlation ($r = -0.20$).  Furthermore, in the presence of strong limb-darkening, the radius ratio and impact parameter are also correlated; the radius ratio can be traded off with the impact parameter to produce similar transit depths.  We point this out because it suggests that the above mentioned discrepancies for $b$ and $R_p/\rs$ are in fact correlated with each other.     

The results for these parameters and other important transit observables are reported in Table 1 (marked with a superscript `a').  Noteworthy are the results for the mean stellar density, $\rho_\star$, and planet surface gravity, $g_p$, which are model-independent determinations making use of information only from transit photometry and Doppler measurements \citep{seager2003,southworth2007,sozzetti2007}.
We also find $a/\rs = 6.23 ~^{+0.71}_{-0.63}$, which is consistent with the determination of W08 ($a/\rs = 7.11 ~^{+0.03}_{-0.81}$).  Note that the corresponding fractional uncertainty in our result for $\rs/a$ is fairly large ($\simeq10\%$).

\section{Stellar and Planetary Properties}

In the transit modeling literature, the parameter of interest, $R_p$, is usually determined via one of the following two methods.  In the first, one obtains an externally determined value of $\rs$, and then multiplies by the light curve results for $R_p/\rs$.  In the second, one assumes a value for $M_\star$, utilizes Newton's version of Kepler's Third Law to derive the semi-major axis, $a$, and then applies the light curve results for $\rs/a$ (and $R_p/\rs$).\footnote{Another possible route is to assume a stellar mass-radius relation \citep{cody2002}.  We have chosen not to make such an assumption because of the uncertainty in the age and evolutionary state of HD 149026.} While the first method has the advantage that transit photometry determines $R_p/\rs$ more precisely than $R_p/a$, the resulting $R_p$ depends strongly on the assumed $\rs$ ($R_p \propto \rs$).  This method also has the disadvantage of effectively disregarding any information gleaned from the light curve on $\rs/a$.   In the second method, the result for $R_p$ depends only weakly on the assumed $M_\star$ ($R_p \propto M_\star^{1/3}$), but, in our case, $R_p/a$ is not constrained well enough to lead to a satisfactorily precise determination of $R_p$.
For these reasons, we adopt a hybrid approach, imposing a radius constraint and, to make use of the $\rs/a$ information, a mass constraint.  Though the addition of this mass constraint represents an increased dependence on stellar models, we consider it a fairly benign dependency given how weakly the mass enters into the transit modeling ($\propto M_\star^{1/3}$).  

In this section, we augment the \emph{Spitzer} dataset with 10 previously published light curves.  Together with independent constraints on the stellar properties described below, we fit for the $R_p$ and other planet and stellar quantities.

\subsection{Stellar Radius and Mass}

Using a collection of interferometric angular diameter measurements, \citet{kervella2004} derived empirical relations for the angular diameters of dwarf stars as a function of Johnson magnitudes.  We use their $V, K$ relation for the angular diameter, $\phi$ (mas),
\begin{equation}
\label{eq:kervella}
\log \phi = 0.0755 (V\!-\!K) + 0.5170 - 0.2 K.
\end{equation}
\citet{kervella2004} find the root mean square residual from this best fit relation to be less than $1\%$ for 20 stars ranging from spectral type A0 to M2.  
We applied $V=8.15$, as found in the \emph{Hipparcos} Catalog, and $K=6.85$, after transforming the 2MASS $K_s$ magnitude to Johnson K following \citet{carpenter2001}. After propagating the uncertainties in the photometry and the Kervella et al. (2004) best-fit parameters, we determined $\phi = 0.1755 \pm 0.0021$ mas (see also Torres et al. 2008).   The formal uncertainty in angular diameter is thus $1 \%$ and negligible compared to the uncertainty in parallax. After combining with the re-reduced \emph{Hipparcos} parallax and uncertainty ($\pi= 12.59 \pm 0.70$) of \citet{vanleeuwen2007}, we determine $\rs = 1.50 \pm 0.09 ~\rsun$.
For the stellar mass, we adopt the value $M_\star= 1.30 \pm 0.10 ~\msun$ from \citet{sato2005}, who derived the value by matching stellar evolution tracks to spectroscopic properties.

\subsection{Light Curve Analysis Revisited}

We simultaneously fitted our \emph{Spitzer} data together with 3 light curves published by \citet{sato2005}, 2 light curves by \citet{charbonneau2006} and 5 light curves by W08.  The 10 previous transit observations are discussed in detail in the references above, but we describe them briefly here.  The Sato et al. 2005 and W08 observations were obtained with 0.8 m automated photometric telescopes at the Fairborn Observatory.  Fluxes were measured simultaneously through Str\"omgren $b$ and $y$ filters and averaged to create $(b\!+\!y)/2$ fluxes.  The Charbonneau et al. (2006) observations were obtained with the Fred Lawrence Whipple Observatory 1.2 m telescope through the Sloan $g$ and $r$ filters.  In analysis of the 10 light curves, W08 divided each raw light curve by a linear function of time that was fitted to the out-of-transit data.  This step corrects for airmass effects and other systematic trends, but also has the effect of normalizing each light curve to have unit mean out-of-transit flux.  We adopted these corrected data as well as their revised photometric errors, which were rescaled to account for the effects of noise correlation on ingress/egress timescales.  Note that the composite of these 10 light curves, when binned to 30 second resolution, shows roughly the same scatter ($\sim 0.9$ mmag) as the the \emph{Spitzer} data binned to the same resolution. 

Next, we revisited the MCMC analysis of \S 3.  We modeled the light curves as before, using the small-planet transit routine of \citet{ma2002}.  For the $g$ and $r$ band data, we assumed linear limb-darkening, with coefficients as tabulated by \citet{claret2004} for a 6250 K, $\log g = 4.5$, and [Fe$/$H]=0.3 star.  For the $(b\!+\!y)/2$ data, we assumed a linear limb-darkening coefficient of 0.712, the average of \citet{claret2000} $b$ and $y$ limb-darkening coefficients, following W08.  We employed 9 free parameters: $R_p/\rs$, $i$, $c_0$, $c_1$, $c_2$, $P$, $T_c$, $\ms$, and $\rs$, where the ramp correction coefficients, $c_i$, apply only to the \emph{Spitzer} data.  Note that we fit for only a single mid-transit time and required the transits to be spaced at integral multiples of $P$.  W08 found no significant deviations from predicted transit times, so this is a reasonable assumption.  We modified our goodness-of-fit statistic as follows:
\begin{equation}
\chi^2 = \sum_{i}{ \left(
\frac{f_{mod}(i)-f_{obs}(i)}{\sigma}\right)^2 } 
+ \left(\frac{\rs/\rsun - 1.50}{0.09}\right)^2
+ \left(\frac{M_\star/\msun - 1.30}{0.10}\right)^2
\end{equation}
with the second and third term reflecting the above determined stellar radius and mass with errors assumed to follow normal distributions.  Note that the mass and radius constraints have entered into the $\chi^2$ in a simple additive form, which is strictly valid only if the constraints were determined entirely independent of each other.  In fact, the mass determination of \citet{sato2005} makes use of the parallax, which implies that the mass and radius determinations have some level of intrinsic covariance.  To examine the impact of this covariance, we repeated our analysis with the following trial goodness-of-fit statistic,
\begin{eqnarray}
\nonumber \chi^2_{trial} & = & \sum_{i}{ \left(
\frac{f_{mod}(i)-f_{obs}(i)}{\sigma}\right)^2 } 
+ \frac{1}{1-\rho_{MR}^2}\left[\left(\frac{\rs/\rsun - 1.50}{0.09}\right)^2
+ \left(\frac{M_\star/\msun - 1.30}{0.10}\right)^2 \right.
\\& & \left.  - 2 \rho_{MR} \left(\frac{\rs/\rsun - 1.50}{0.09}\right)\left(\frac{M_\star/\msun - 1.30}{0.10}\right) \right]
\end{eqnarray}
where we experimented with values of the correlation coefficient, $\rho_{MR}$, between -1 and 1.  For $-0.7 < \rho_{MR}< 0.7$, we found the best-fit parameters and error bars to be negligibly affected by the correlation, and we found the results were significantly impacted only when $|\rho_{MR}| > 0.9$.  Since we expect the covariance between the stellar mass and radius determination to be more modest, we conclude that our results are not impacted by neglecting the covariance. 

We conducted the analysis as before; we produced 10 Monte Carlo chains of length $10^6$, cut the first $25\%$ of each chain, and then combined the chains.  For each parameter the Gelman-Rubin R statistic was well within $1 \%$ of unity.  In table 1, we report best-fit values and uncertainties for various parameters.  We take the best-fit value to be the median of the MCMC samples, and for the uncertainties, we report the interval that encloses the central $68.3\%$ of the MCMC samples.

We determine a stellar radius of $1.497 \pm 0.069 ~\rsun$, which is moderately refined compared to its prior distribution ($1.50 \pm 0.09 ~\rsun$).  This refinement indicates that the combination of \emph{Spitzer} and ground-based data is able provide some statistical influence on the parameter estimation through the observational constraint on $\rs/a$.  
We note that the external stellar radius constraint reinforces the high impact parameter solutions favored by the analysis of \S 3 (\emph{Spitzer} data alone).  This arises because the radius constraint favors relatively large stellar radii, which, for the given observed transit duration, can only be accommodated by non-equatorial impact parameters. 
The planet radius is determined to be $0.755 \pm 0.040 ~\rjup$, with an uncertainty that is reduced versus previously published determinations (for example, $0.71 \pm 0.05 ~\rjup$ as determined by W08).  The reduction is partly due to the smaller uncertainty in the revised \emph{Hipparcos} parallax \citep{vanleeuwen2007} that we have adopted, and partly due to the combination of the \emph{Spitzer} and ground-based data.  


\clearpage

\begin{figure} 
\epsscale{1.0}
\plotone{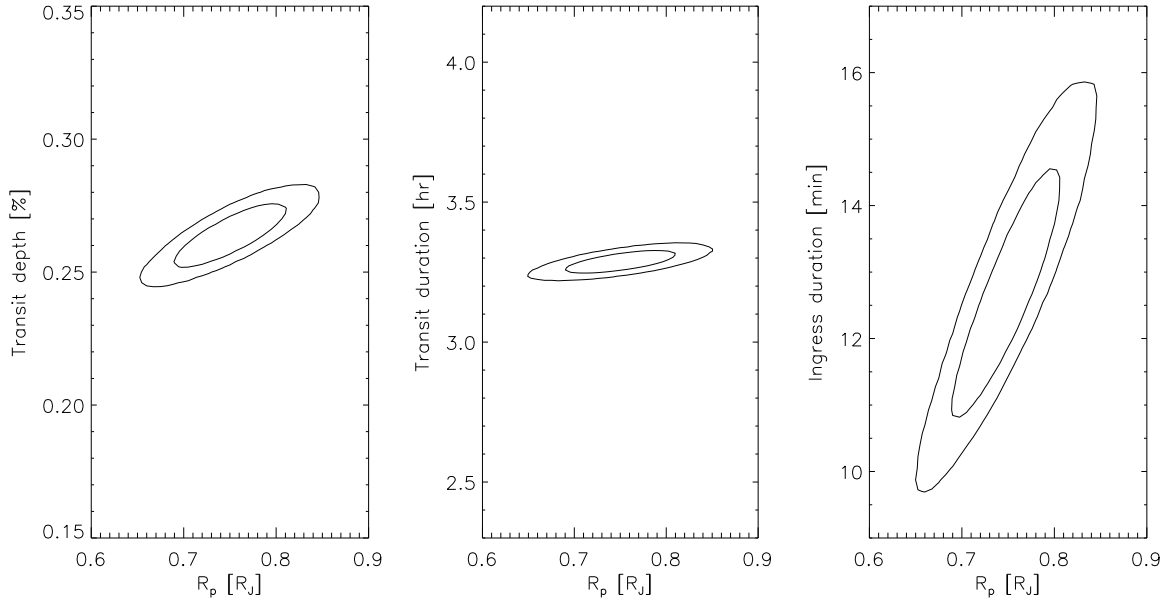}
\caption{Joint posterior probability distributions for $R_p$ and directly observable quantities, as estimated by the MCMC analysis described in \S 4.  
The contours mark the $68\%$ and $95\%$ confindence regions. Left: Joint distribution for $R_p$ and the ``transit depth,'' $(R_p/\rs)^2$.  
Middle: Joint distribution for $R_p$ and the transit duration, defined as the interval from 1$^{st}$ to 4$^{th}$ contact.  
Right: Joint distribution for $R_p$ and the ingress (egress) duration, defined as the interval from 1$^{st}$ to 2$^{nd}$ (3$^{rd}$ to 4$^{th}$) contact.  Note that the vertical axes in each panel are scaled so that they encompass roughly the same fractional variation. } 

\label{histos2}
\end{figure}

\clearpage

In Fig.~\ref{histos2}, we look deeper into the observational constraints on the key parameter, $R_p$.  The most directly observable quantities from a transit light curve are the transit depth, the total transit duration, and the ingress or egress duration; a measurement of these three observables is sufficient (at least in the absence of limb-darkening) to determine the more physical parameters $R_p/\rs$, $\rs/a$, and $\cos i$.  By examining the joint posterior distributions for the three observables with the parameter $R_p$, one can gain insight into the current observational limitations on the precision of $R_p$.  While each of the panels in Fig. \ref{histos2} demonstrate covariances, the third panel ($R_p$ and ingress duration) shows particularly strong covariance.  Thus, the major limiting factor in reducing the uncertainty in planetary radius appears to be the ability to resolve the ingress duration.  Unfortunately, constraining this quantity with ground-based photometry is complicated by limb-darkening and the effects of systematics and correlated errors \citep[e.g.,][]{pont2007}. 
  

\subsection{Influence of Star Spots}

As with limb-darkening, an inhomogeneous surface brightness due to spots would impact both the depth and shape of the transit light curve.  If the transit chord intersects a star spot, a positive ``bump'' will be introduced into the transit light curve, while if the transit chord is along an unspotted area of an otherwise spotted star, the transit would appear deeper (see e.g. Knutson et al. 2008, Beaulieu et al. 2008).  The existence of star spots can be investigated by long-term photometric observations of the star, monitoring for periodic flux variations.

Previously published APT data has shown HD 149026 to be photometrically stable to 0.0015 mag, the limit of precision of the APTs (Sato et al. 2005).  Further $(b+y)/2$ out-of-transit observations have been obtained with the APTs, extending the dataset to over 3 years (Winn et al. 2008; G. Henry, private communication).  With this additional APT data, kindly shared with us by G. Henry, we have searched for evidence of star spot-induced variability.
We computed the periodogram for the time-series (550 total flux measurements) in fine steps of the period for periods between 0.5 and 100 days.  Examination of all prominent peaks in the periodogram reveals no evidence for any significant periodicities, and allows us to place an upper limit on the peak-to-peak amplitude of any sinusoid (in the period range 0.5-100 days) of less than 0.001 mag.  Any spots at or below this level will have negligible impact on the transit light curve, especially given that the 1 $\sigma$ uncertainty in the transit depth for HD 149026b is $3\%$.

\subsection{Refined Ephemeris}
\begin{figure} 
\epsscale{1.0}
\plotone{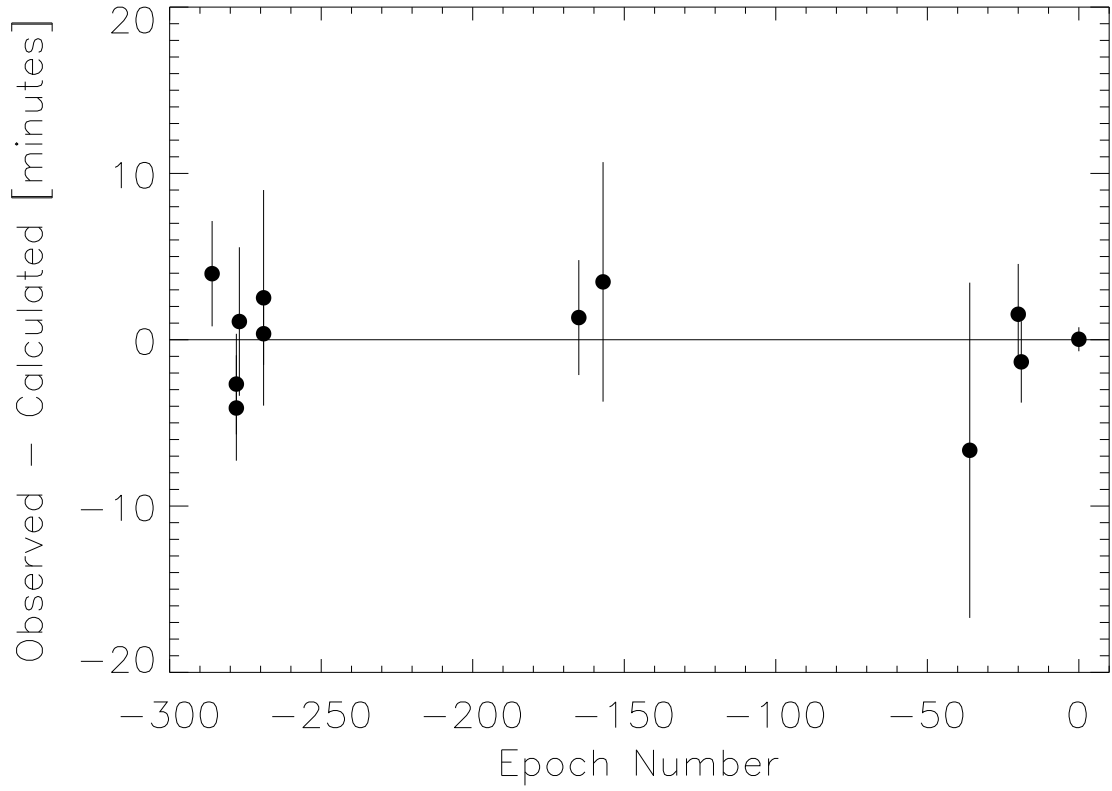}
\caption{Observed minus calculated mid-transit times for HD 149026b.  The calculated transit times are derived from the ephemeris in eq. (2). The estimates for the first 11 transit times are drawn from Table 3 of W08.} 
\label{oc}
\end{figure}

The precise transit timing from \S 3, along with the fact that the \emph{Spitzer} light curve extends the time base-line of HD 149026b transit observations, enables a significant refinement in the transit ephemeris.  For the previously published transit observations, we adopt the transit times and uncertainties listed in Table 3 of W08.  We fit the timing data to the equation
\begin{equation}
T_c(E)= T_c(0) + E \times P
\end{equation}
where $T_c$ is the transit time, $E$ is the transit epoch, and $P$ is the orbital period.  We determine $P=2.8758887 \pm 0.0000035$ and $T_c(0)=2454327.37211 \pm 0.00047$, with $\chi^2/N_{\rm{dof}} = 0.564$, with $N_{\rm{dof}} = 10$.  In Fig. \ref{oc}, we show the transit time residuals for all published transits.

\clearpage

\clearpage

\section{Discussion}

We have presented and analyzed \emph{Spitzer} 8 $\mu$m transit observations of the HD 149026 system.  By incorporating previously published data, and adopting constraints on the stellar mass and radius, we improve the determination of the planetary radius to $R_p = 0.755 \pm 0.040 ~\rjup$.  Our measurement reinforces previous findings of the intriguingly small radius of HD 149026b.  To place this result in context, models in which HD 149026b (with a total mass of 114 $\pm 2 ~\mearth$ ) is composed purely of H/He require a radius greater than $1.1 ~\rjup$ \citep[e.g.,][]{burrows2007}.

The implications of the small measured radius on the interior structure of HD 149026b have been modeled by a number of authors.  Most works \citep{sato2005,fortney2006,ikoma2006,burrows2007} have assumed that all of the heavy elements reside within the planet's core, although it was often stressed that this may not necessarily be the case.  For instance, recent models of Jupiter's structure indicate that the majority of its heavy elements are mixed within the H/He envelope (Saumon \& Guillot 2004).  \citet{baraffe2008} recently computed evolution models of HD 149026b and other planets and showed that if these heavy elements are distributed within the envelope, rather than all in the core, less are needed to obtain the same model radius at a given age.  However, \citet{ikoma2006} also explored this effect, and noted that enhanced metallicity of the H/He envelope should also lead to higher atmospheric opacity, which will slow the contraction, necessitating more heavy elements.

The choices made by the modelers have been diverse, and many different atmospheric boundary conditions, assumed heat capacities of the heavy elements, and equations of state (EOSs) for the heavy elements have been explored.  Interior heavy element mass estimates have generally ranged from 50-90 $\mearth$ for a planet radius of $0.725 \pm 0.05 R_J$.  On the low end, \citet{ikoma2006} find that a $35 \mearth$ core would be necessary, if the planet cooled and contracted in isolation, then was brought to 0.042 AU at the present time.  On the high end, \citet{burrows2007} find 110 $\mearth$, if the planet has an atmospheric opacity 10 times larger than solar composition atmosphere models.  For all of these models, uncertainty in the measured radius is more significant than the uncertainty in the system age.

A full exploration of new evolution models, including the potential contribution of TiO/VO opacity, which may be present in the planet's visible atmosphere (Fortney et al. 2006, Fortney et al. 2008, Burrows et al. 2008), is beyond the scope of this paper.  Given the previous modeling efforts, together with uncertainties in atmospheric metallicity and opacities and the distribution of heavy elements within the planet, the 50-90 $\mearth$ heavy element mass range is still likely to be correct, even for our modestly larger measured radius.  We note that current estimates of the heavy element abundance of Saturn (which is similar in mass to HD 149026b) and Jupiter range from 13-28 $\mearth$ and 8-39 $\mearth$, respectively \citep{saumon2004}.  Uncertainty in the composition of giant planets is the rule, not the exception.

As has been stressed recently by \citet{burrows2007}, and others, constraints for any particular planet will remain uncertain, but with a large sample size of transiting planets at various masses, radii, orbital distances, and stellar metallicity, trends will emerge which will shed light on the formation and structure of these planets \citep{guillot2006,fortney2007,burrows2007}. 

We are thankful to F. van Leeuwen for providing parallax data for HD 149026, and to G. Takeda for discussions regarding the spectroscopic determination of stellar properties.  We are especially grateful to G. Henry for sharing many seasons of photometric data. We would also like to thank an anonymous referee for specific and helpful recommendations.  This work is based on observations made with the Spitzer Space Telescope, which is operated by the Jet Propulsion Laboratory, California Institute of Technology under a contract with NASA. Support for this work was provided by NASA through an award issued by JPL/Caltech.

\bibliographystyle{apj}

\begin{deluxetable}{llll}
\tabletypesize{\normalsize}
\tablecaption{Estimates of the HD~149026 System Parameters}
\tablewidth{0pt}
\label{posteriori}
\tablehead{
  \colhead{Parameter} & \colhead{Median} & \colhead{15.9$^{th}$ Percentile} & \colhead{84.1$^{st}$ Percentile}
}

\startdata
$R_p/\rs$\tablenotemark{a}	& 0.05158	& $-$0.00077	& $+$0.00077	\\
$R_p/\rs$			& 0.05147	& $-$0.00077	& $+$0.00076	\\
$i$~[deg]\tablenotemark{a}     	& $ 85 \fdg4 $ 	& $-1 \fdg 9 $ 	& $+2 \fdg 5$  	\\
$i$~[deg]			& $ 85 \fdg3 $	& $-0 \fdg 8 $	& $+0 \fdg 9$	\\
$a/\rs$\tablenotemark{a}	& 6.23		& $-$0.63		& $+$0.71		\\
$a/\rs$				& 6.20		& $-$0.25		& $+$0.28		\\
$\rho_\star$[g cm$^{-3}$]\tablenotemark{a}& 0.51 & $-$0.13		& $+$0.21		\\
$\rho_\star$[g cm$^{-3}$]	& 0.547		& $-$0.064	& $+$0.076	\\
$\log g_p$[cgs]\tablenotemark{a,b} &  3.18 	& $-$0.09 	& $+$0.10		\\
$\log g_p$[cgs]\tablenotemark{b}& 3.203		& $-$0.048	& $+$0.049	\\
$P$ [days]			&2.8758887 	& $-$0.0000035	& $+$0.0000035 \\
$T_c$ [HJD]\tablenotemark{a}	&2454327.37213	& $-$0.00050	& $+$0.00050	\\
$R_p [\rjup]$ 			& 0.755  	& $-$0.040 	& $+$0.040 	\\
$\rs  [\rsun]$  	        & 1.497   	& $-$0.069   	& $+$0.069 	\\
\enddata

\tablenotetext{a}{Determined from analysis of \emph{Spitzer} data alone}
\tablenotetext{b}{Using $K=43.3 \pm 1.2$~m~s$^{-1}$, from Sato et
  al.~(2005)}
\end{deluxetable}

\end{document}